\documentclass[12pt,preprint]{aastex}

\shorttitle{X--ray/Optical follow-up of PKS~2155--304}
\shortauthors{Foschini et al.}

\begin{document}

\title{X--ray/UV/Optical follow-up of the blazar PKS 2155--304 after 
the giant TeV flares of July 2006}


\author{L. Foschini\altaffilmark{1,*}, G. Ghisellini\altaffilmark{2}, 
F. Tavecchio\altaffilmark{2}, A. Treves\altaffilmark{3},
L. Maraschi\altaffilmark{2}, M. Gliozzi\altaffilmark{4}, C.M. Raiteri\altaffilmark{5}, 
M. Villata\altaffilmark{5}, E. Pian\altaffilmark{6}, G. Tagliaferri\altaffilmark{2}, 
G. Tosti\altaffilmark{7}, 
R.M. Sambruna\altaffilmark{8}, G. Malaguti\altaffilmark{1},  G. Di Cocco\altaffilmark{1}, 
P. Giommi\altaffilmark{9}}

\altaffiltext{1}{INAF/IASF-Bologna, Via Gobetti 101, 40129 Bologna (Italy)}
\altaffiltext{2}{INAF/Osservatorio Astronomico di Brera, Via Brera 28, 20121 Milano (Italy)}
\altaffiltext{3}{Dipartimento di Scienze, Universit\`a dell'Insubria, 
Via Vallegio 11, 22100 Como (Italy)}
\altaffiltext{4}{George Mason University, Department of Physics and Astronomy, School 
of Computational Sciences, Mail Stop 3F3, 4400 University Drive,
Fairfax, VA 22030, USA}
\altaffiltext{5}{INAF/Osservatorio Astronomico di Torino, Via Osservatorio 20, 10025 
Pino Torinese (Italy)}
\altaffiltext{6}{INAF/Osservatorio Astronomico di Trieste, Via G.B. 
Tiepolo 11, 34131 Trieste (Italy)}
\altaffiltext{7}{Osservatorio Astronomico, Universit\`a di Perugia, Via B. Bonfigli, 
06126 Perugia (Italy)}
\altaffiltext{8}{NASA/Goddard Space Flight Center, Code 661, Greenbelt, MD 20771, USA}
\altaffiltext{9}{ASI Science Data Centre, Via G. Galilei, 00044 Frascati, (Italy) }

\altaffiltext{*}{Email: \texttt{foschini@iasfbo.inaf.it}}

\begin{abstract}
We present  all the publicly available data, from optical/UV 
wavelengths (UVOT) to X--rays  (XRT, BAT), obtained from
\emph{Swift} observations of the blazar PKS~2155--304, 
performed in response to the rapid alert sent out
after the strong TeV activity (up to $17$ Crab flux level at
$E>200$~GeV) at the end of July 2006. The X-ray flux
increased by a factor of 5 in the $0.3-10$~keV energy band and by a
factor of $1.5$ at optical/UV wavelengths, with roughly one day of
delay. The comparison of the spectral energy distribution built with
data quasi-simultaneous to the TeV detections shows an increase of the
overall normalization with respect to archival data, but only  a
small shift of the frequency of the synchrotron peak that remains
consistent with the values reported in past observations when the
TeV activity was much weaker.

\end{abstract}


\section{Introduction}

PKS~2155--304 ($z=0.116$) is one of the best known blazars and the
second brightest in X-rays (after Mkn 421),  observed
many times at various  wavelengths. At $\gamma-$ray energies, it
was first detected  by CGRO/EGRET with a photon index
$\Gamma=1.71\pm 0.24$ in the energy range $0.03-10$~GeV (Vestrand et
al. 1995). This hard spectrum suggested a possible detection in the
TeV energy range (Vestrand et al. 1995, Stecker et al. 1996, Tavecchio
et al. 1998) that was first achieved by the University of Durham Mark
6 $\check{\rm C}$erenkov telescope in $1996-1997$ (Chadwick et
al. 1999).

In 2002--2003 the TeV activity has been monitored by
HESS\footnote{\texttt{http://www.mpi-hd.mpg.de/hfm/HESS/}} that
detected a flux variation in the $E>300$~GeV energy band from $1.2$ to
$7.8\times 10^{-11}$~ph~cm$^{-2}$~s$^{-1}$ equivalent to $90-560$
mCrab (Aharonian et al. 2005a). At the end of July 2006, the source
displayed an anomalously high activity. Preliminary analysis of the HESS
data showed an average flux level of $8$~Crab in the $E>200$~GeV
energy band with flares up to $17$~Crab during the night of
$27-28$~July, an average flux level of $1$~Crab during the night of
$28-29$~July with smaller activity, while a second outburst occurred 
in the night of $29-30$~July with an average of $5$~Crab and flares up 
to $13$~Crab (Raue et al. 2006, Aharonian et al., in prep.). 
The event was detected also by
MAGIC\footnote{\texttt{http://wwwmagic.mppmu.mpg.de/}} (A. De Angelis,
private communication). Following a rapid alert (Benbow et al. 2006)
some high-energy satellites pointed at PKS~2155--304.

The \emph{Swift} satellite (Gehrels et al. 2004) performed optical/UV
and X-ray follow-up, starting on July~$29$ ending on August~$29$
2006, with repeated short exposure pointings. Here we present the results
of these observations, a comparison with previous ones, when PKS~2155--304 
was in a low activity state and theoretical modeling of the SEDs.

\section{Data analysis}

The data from the three instruments onboard \emph{Swift} have been
processed and analyzed with \texttt{HEASoft v. 6.1.2} with the latest
calibration files (6 December 2006). 

Data from individual pointings
from the coded-mask hard X-rays detector BAT (optimized for the
$15-150$~keV energy band, Barthelmy et al. 2005) were binned, cleaned
from hot pixels and background, and deconvolved. The intensity images
were then integrated by using the variance as weighting
factor. PKS~2155--304 was not detected either in individual pointings
nor in the integrated mosaic image. The upper limit for a $3\sigma$
detection in the $20-40$~keV energy band -- already corrected for
systematics -- is $3.3\times 10^{-10}$~erg~cm$^{-2}$~s$^{-1}$
($42$~mCrab) for the pointing of $29$~July~$2006$ (exposure $6$~ks)
and $1.6\times 10^{-10}$~erg~cm$^{-2}$~s$^{-1}$ ($20$~mCrab) for the
overall mosaic (exposure $30$~ks).

\begin{figure}
\centering
\includegraphics[angle=270,scale=0.6]{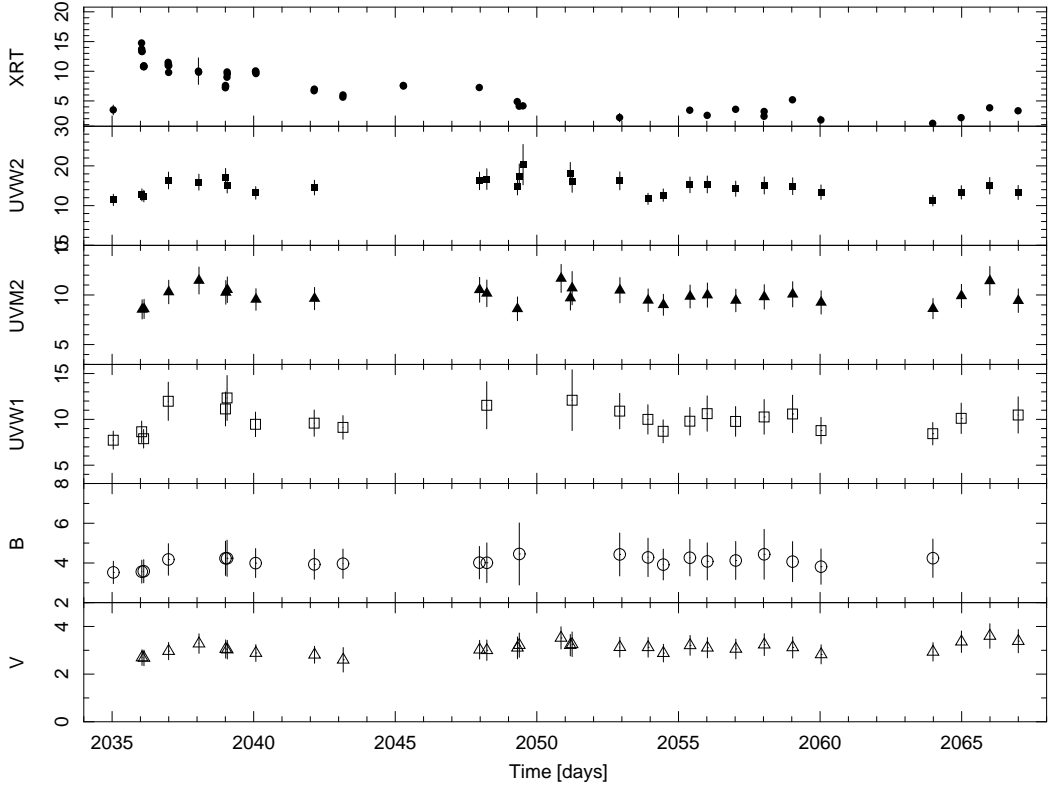}
\caption{Lightcurves built from data of \emph{Swift} instruments. From top to bottom: 
XRT [$0.3-10$~keV], UVOT UVW2 ($1880$~\AA), UVM2 ($2170$~\AA), UVW1 ($2510$~\AA), 
B ($4390$~\AA), V ($5440$~\AA). For XRT: only WT mode data have been included, binned to
$500$~s; a flux of $10^{-10}$~erg~cm$^{-2}$~s$^{-1}$ is approximately equal to $3.5-4.5$~c/s. For UVOT: 
U filter data are not included, since for most of the observing time the detector was 
saturated; for all the other filters, the flux is given in units 
$10^{-14}$~erg~cm$^{-2}$~s$^{-1}$~\AA$^{-1}$ and is not corrected for absorption. 
The mark on abscissa indicates the $00:00:00$~UTC of the day, i.e. the
day $2035$ corresponds to $29$ July $2006$ at $00:00:00$. 
\label{fig:swiftlc}}
\end{figure}

Data from the X-ray Telescope XRT ($0.3-10$~keV, Burrows et al. 2005)
were analyzed  using the \texttt{xrtpipeline} task. XRT
automatically switches the operating mode according to the target
source flux, changing from window timing (WT, high flux) to photon
counting (PC, low flux) with a threshold around $1$~mCrab ($5\times
10^{-11}$~erg~cm$^{-2}$~s$^{-1}$). The X-ray flux of PKS~2155--304
remained almost always above $\approx 10^{-10}$~erg~cm$^{-2}$~s$^{-1}$
(Fig.~\ref{fig:swiftlc}) and therefore we analyzed only the window
timing mode data. There are also a few hundreds of seconds exposure in
photon counting mode, but the point-spread function (PSF) is severely
affected by pile-up. Nevertheless, we used the source position
measured in the images accumulated with PC data as best input for the
pipeline of WT mode (without imaging). We selected only the grade 0
(single pixel) events and extracted the spectra only from pointings
with exposures greater than $100$~s, in order to have the best
available statistics. The remaining pointings (i.e. with less than
$100$~s) are anyway included in the lightcurve displayed in the top panel
of Fig.~\ref{fig:swiftlc}, in order to give a better coverage of the time
evolution of the source.

Since most of the pointings lasted a
few hundreds of seconds, there exposure is insufficient to have data
at energies above $4-6$~keV, except for the pointing of 29 July, where
a $5$~ks exposure allowed us to have useful signal up to $8$~keV.  In
addition, the XRT response is limited at low energies, because there
are still some residual instrumental feature around $0.5$~keV (Campana
et al. 2006). Therefore, the extracted spectra were fitted in the
range $0.3-0.45$~keV and from $0.6$ to $4-8$~keV, depending on the
statistics and then the flux was measured in the full $0.3-10$~keV
band. The results are summarized in Table~\ref{table:swiftlog}.

For XRT, we note that the observation indicated in Table~\ref{table:swiftlog} as ``29/31-07'' 
actually started on $29$~July~$2006$ at $00:55:42$~UTC and ended on $31$~July~$2006$ at
$00:01:00$~UTC, resulting in an elapsed time of $\approx 1.7\times 10^{5}$~s. However, the 
effective exposure time is only $4916$~s. Thus, the spectral data reported in Table~\ref{table:swiftlog}
refer to the average of the snapshot observations during this period. In addition, we
also extracted from this observation, the subset of data referring only to the night
between 29 and 30 July, in order study the data available that are simultaneous to TeV observations.
The spectral information about that night are indicated in Table~\ref{table:swiftlog} as
``30-07''.

Data from the optical/UV telescope UVOT (Roming et al. 2005) were
analyzed by using the \texttt{uvotmaghist} task with a source region
of $6''$ for optical and $12''$ for UV filters. The background was
extracted from an annular region centered on the source and with an
inner region equal to the source region plus $2''$ and the outer
radius equal to $60''$.  To take into account systematic effects, we
added a $10\%$ error in flux (resulting in about
$0.1$~magnitudes). The results, simultaneous to the X-ray fits, are
summarized in Table~\ref{table:swiftlog}, while complete lightcurves
are shown in Fig.~\ref{fig:swiftlc}.

In order to compare X-rays/UV/optical data close to the outburst with
data when PKS~2155--304 was not active 
(i.e. with low X-ray/UV/optical fluxes), we retrieved and 
analyzed \emph{Swift} observations of the blazar performed in April $2006$.
The results of the analysis are reported in Table~\ref{table:swiftlog}
where the flux difference between April and August observations
shows up clearly.

\section{Discussion and interpretation}

\subsection{Overview of data}

\begin{figure}
\centering
\includegraphics[angle=270,scale=0.6]{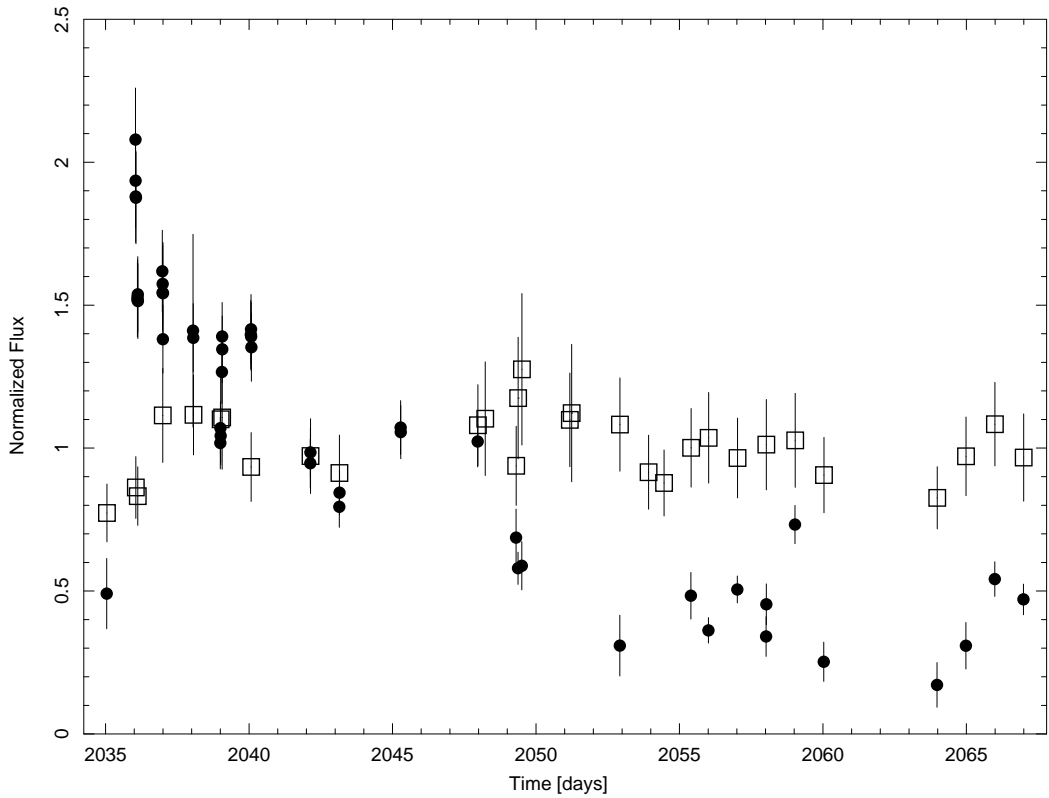}
\caption{Normalized lightcurves. Black filled circles indicate XRT data [$0.3-10$~keV] 
normalized to their average in July-August $2006$ ($7.1\pm 0.6$ c/s). 
Open squares indicate the average from UV filters, which are UVW2 ($1880$~\AA), 
UVM2 ($2170$~\AA), and UVW1 ($2510$~\AA), normalized to 
$(11.6\pm 0.6)\times 10^{-14}$~erg~cm$^{-2}$~s$^{-1}$~\AA$^{-1}$. 
\label{fig:swiftlc2}}
\end{figure}

The \emph{Swift} observations of PKS~$2155-304$ starting on July 29th at $00:55$
during the phase of strong TeV activity reported by HESS (Raue et
al. 2006, Aharonian et al., in prep.), show an initial
increase of the X-ray flux, by a factor 4, between the observation of
29 July and that of 30 July followed by an overall decrease, while
optical/UV fluxes show a moderate activity
(Fig. \ref{fig:swiftlc2}). No detection in hard X-rays was obtained
with BAT.

With respect to the April 2006 observations, XRT recorded a
change by a factor $5$, while the UV flux increased by a factor of $\approx 1.5$. For comparison 
the HESS observations of July 2006 showed "night-averaged" intensities in
the TeV band of factors $\approx 16$ and $\approx 10$ larger than those 
in $2002-2003$ ($0.5$~Crab; Aharonian et al. 2005),
but with short flares of up to a factor of $34$.

The sparseness of the available data does not allow us to make
stringent correlations with TeV data. We note however that the initial
flare in X-rays taking place between the nights of 29 and 30 July, approximately 
coincides with the second TeV outburst, while a lower amplitude flare in the UV 
occurs about one day later (Fig.~\ref{fig:swiftlc2}). Since the UVOT detector often saturated
because of coincidence losses, we cannot exclude the occurrence of
other flares with greater amplitude.


\subsection{Spectral Energy Distribution}

\begin{figure}
\centering
\includegraphics[scale=0.6]{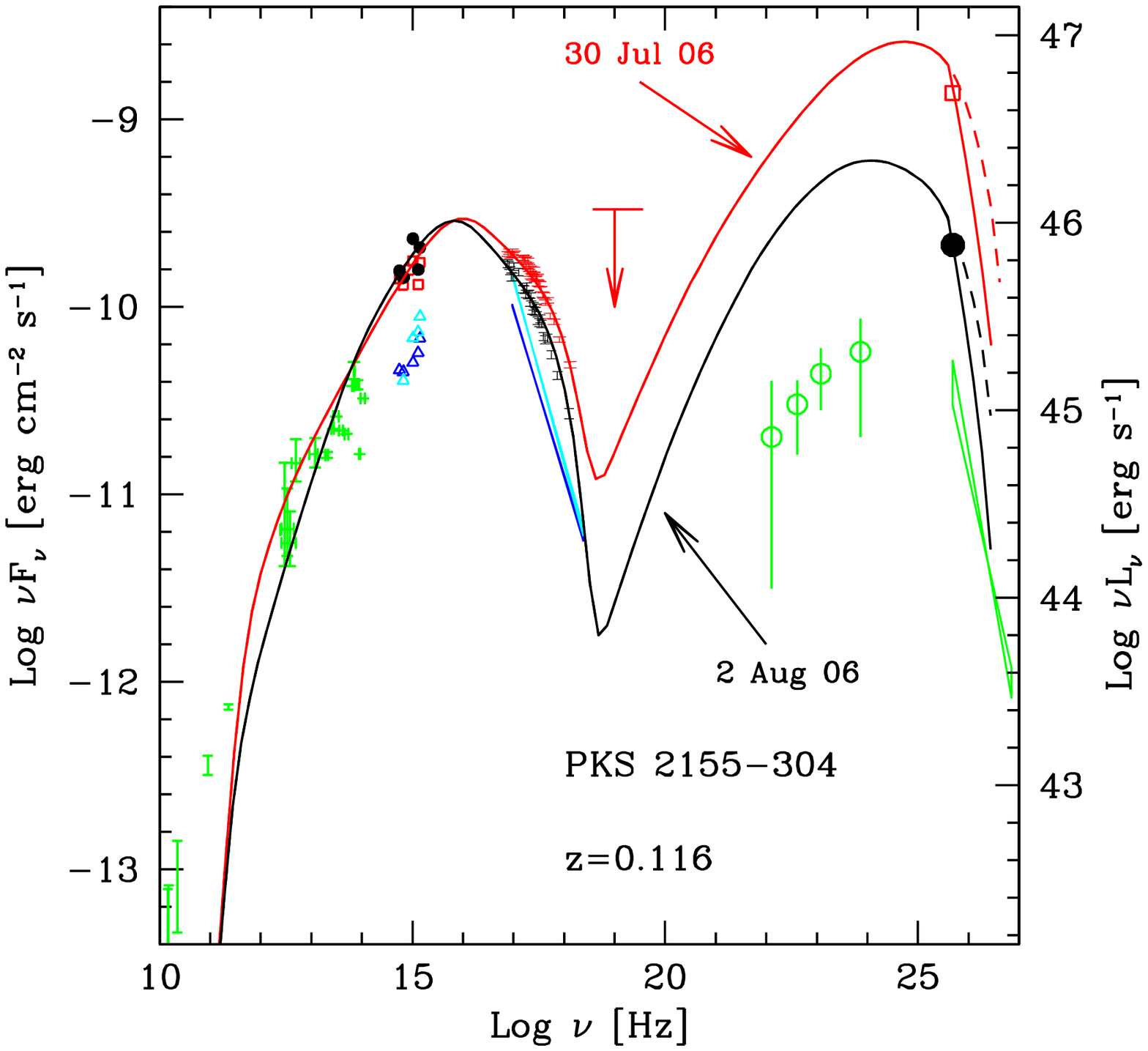}
\caption{Spectral Energy Distribution of PKS~2155--304: the red
symbols are the quasi-simultaneous data of 29 July; the black symbols
refer to the observations of 2 August. HESS data from Raue et
al. 2006; \emph{Swift} quasi-simultaneous data are from the present
work. For comparison, we also report data from the historical
records: green symbols refer to archival data (references in
Chiappetti et al. 1999) and to the H.E.S.S. TeV spectrum taken in
October-November 2003 (Aharonian et al. 2005b), while other colors
report the \emph{XMM-Newton} data from Foschini et al. (2006). The red
continuous line represents the synchrotron self Compton model (SSC,
see Ghisellini et al. 2002) used to fit the data of 29 July 2006,
while the black line represents the model fitted to the data of 2
August. Both models include the absorption at TeV energies due to the
extragalactic infrared background calculated according to Stecker \&
Scully (2006).  The dashed line indicates the instrinsic (i.e. not
absorbed) spectrum.\label{fig:sed}}
\end{figure}

With the current data set we cannot probe the short (5 min) timescale
variability preliminarily reported by HESS (Raue et al. 2006;
Aharonian et al., in prep.), but can investigate only quantities
averaged over timescales of days.  We therefore assembled the spectral
energy distribution (SED, Fig.~\ref{fig:sed}) using the \emph{Swift}
observation tagged as ``30 July'' that is quasi-simultaneous to the second
TeV flare occurred during the night between $29-30$ July 2006.  
We also considered TeV and \emph{Swift} observations performed on 2 August, 
when the blazar activity was declining.

In order to discuss the observed SEDs in terms of changes of relevant
physical quantities, we used the model by Ghisellini et al. (2002) to
reproduce the SEDs in Fig.~\ref{fig:sed}.  As generally assumed for
this and the other TeV BL Lacs (e.g. Aharonian 2004) the X-ray
emission is attributed to synchrotron radiation and the $\gamma$--ray
component to the synchrotron self-Compton (SSC) process.  The source
is assumed to be a sphere of radius $R$ travelling with bulk Lorentz
factor $\Gamma$ at an angle $\theta$ with respect to the line of
sight, yielding a Doppler factor $\delta$. The magnetic field $B$ is
tangled and homogeneous.  The distribution of emitting relativistic
electrons is computed as the result of a broken power law injection
distribution $\propto \gamma^{-s}$ between $\gamma_1$ and $\gamma_2$,
and $\propto \gamma^{-1}$ below $\gamma_1$, subject to radiative
cooling occurring in a light crossing time $R/c$.  This injection of
relativistic particles correspond to an injected power $L_{\rm inj}'$
as measured in the comoving frame.  The resulting particle
distribution $N(\gamma)$ is formed by power law segments, the steepest
of which is $\propto \gamma^{-(s+1)}$.
%
%
All the parameters corresponding to the two models 
shown in Fig.~\ref{fig:sed} are listed in Table~\ref{param}. 
Tavecchio et al. (1998) have shown that, in principle, if the 
peak frequencies and fluxes of the synchrotron and self--Compton
components and the variability timescales are known, then the 
main parameters for the SSC model are determined.
At present we have only a very partial knowledge of the SED at high energies,
resulting in ambiguities in the parameter choice.
To fix them, we minimized the luminosity in the
self--Compton component assuming an intrinsically steep TeV spectrum.
We have also assumed that the variability timescales is 
$\sim$1 hour, as typically observed in the X--ray band
(e.g. Zhang et al. 2002).
The much shorter 5 minute variability timescale recently observed in the
TeV band should then imply a different additional region/emission process.

The model results show that, while there is some evidence of a flatter X-ray spectrum
at higher intensity, the frequency of the sychrotron peak remains at $\approx
10^{15-16}$~Hz, consistent with  other observations
with much weaker TeV activity (cf Urry et al. 1997, Chiappetti et
al. 1999, Foschini et al. 2006). The low peak frequency value is also a result
of the choice of reproducing the optical-UV fluxes.
The large difference in TeV fluxes  associated with  small 
differences in X-ray spectra requires, in SSC models,  
an increase of the relativistic electrons accompanied by a decrease of the 
magnetic field. 

The above point is confirmed  by a comparison of the present data and model 
with the \emph{BeppoSAX} observation in November $1997$ 
(Chiappetti et al. 1999) which was
performed quasi-simultaneously to the TeV observations by Chadwick et
al. (1999), when PKS~2155--304 was at about $0.3$~Crab (average flux,
$E>300$~GeV).
With respect to the parameters derived for the November 1997 episode 
the present SSC models yield (cf. Table~\ref{param}) 
a  larger Doppler factor ($\delta=33$ vs 18), a smaller magnetic field
($B=$0.27--0.55 G vs 1 G), a flatter index of the electron distribution 
($p=s+1=$ 3.5--3.6 vs 4.85), and a smaller frequency of the synchrotron peak 
($\approx 10^{16}$ vs $10^{17}$~Hz) with very similar emitting regions
($R=5\times 10^{15}$ vs $3\times 10^{15}$~cm).

In summary, within a simple SSC scheme, the physical parameters of the source 
changed, in the sense of a harder particle spectrum,
a smaller magnetic field and a greater beaming factor in the 2006
observations. 
This is required by the the different self--Compton to synchrotron
luminosity ratio, which was substantially larger in the 2006 observations.

\subsection{Comparison with other cases: Mkn 501 in 1997, Mkn 421 1998-2000}

It is interesting to compare the present episode with other
exceptional activity states occurred in the past in blazar sources
with similar SEDs (HBL, Padovani \& Giommi 1995). The most striking
example todate is the strong TeV activity exhibited by Mkn 501 in
April 1997, observed by the Whipple Observatory: during the nights
from 7 to 19 April 1997, its flux ($E>350$~GeV) changed from $0.5$ to
the peak of $3.8$~Crab occurred on 16 April, with an average of
$1.6$~Crab and no hourly timescale variability (Catanese et
al. 1997). \emph{BeppoSAX} observed the simultaneous highly chromatic
evolution of the source in X-rays: the flux increased by factors of
$4.2$, $2.4$, and $1.5$ in the $13-200$~keV, $2-10$~keV, and
$0.1-2$~keV energy bands respectively, resulting in a frequency shift
of the synchrotron peak by two orders of magnitude (Pian et al. 1998,
Tavecchio et al. 2001). RXTE observations revealed also timescales
variability down to a few tens of minutes (Xue \& Cui
2005). Observations with U filter showed a modest increase of 1\% in
flux (Catanese et al. 1997).  A less extreme, though analogous,
behaviour was observed in Mkn 421 in 1998-2000.  The X-ray and TeV
activity were correlated also on short timescale (Maraschi et al 1999,
Takahashi et al. 2000) with larger amplitude variations in the TeV
band. The synchrotron peak appeared to shift to higher energies but
not as dramatically as for Mkn 501.

The behaviour of PKS~2155-304 appears less striking in X-rays than for
the previous two sources but more extreme in the TeV variability.  The
important questions to be answered concern the understanding of these
different "modes" of variability in terms of physical models of the
sources.  The upcoming gamma-ray missions (AGILE -GLAST) and the
continuous developments of Cherenkov Telescope facilities will allow
to define the spectral variability at high energies with unprecedented
accuracy. It is however mandatory to complement the high energy data
with extensive observations in the X-ray band in order to approach the
physical origin of the variability.

\section{Conclusions}

We presented the observations of the blazar PKS~2155--304  performed by 
the \emph{Swift} satellite immediately after the giant TeV flare observed by HESS at
the end of July 2006 (Raue et al. 2006; Aharonian et al., in prep.).
The most important result appears to be that, in correspondence with the dramatic TeV
activity, the X-ray intensity changed by a factor 5 but without large spectral
changes. In particular the frequency of the synchrotron peak remained at values 
similar to those observed in the past (e.g. 1997, Chiappetti et al. 1999), during low
TeV activity. Modeling of the SED based on the SSC process in a homogeneous 
region suggests  an increase of the Doppler factor ($33$ in $2006$; $18$ in $1997$) 
and of the relativistic electrons associated with a decrease of the magnetic field
($0.27$ G in $2006$; $1$ G in $1997$).


\acknowledgements LF thanks V. Bianchin for useful discussions. This
research has made use of data obtained from the High Energy
Astrophysics Science Archive Research Center (HEASARC), provided by
NASA's Goddard Space Flight Center.

\clearpage

\begin{deluxetable}{cccccccccccc}
\tabletypesize{\tiny}
\tablecaption{\footnotesize{Summary for Swift observations. See Fig.~\ref{fig:swiftlc} for the complete set of data.}\label{table:swiftlog}}
\tablewidth{0pt}
\tablehead{
\colhead{Date} & \colhead{XRT Exposure [s]} & \colhead{Parameters\tablenotemark{a}} 
& \colhead{F\tablenotemark{b}} & \colhead{$\tilde{\chi}^2$/dof} &
\colhead{V\tablenotemark{c}} & \colhead{B\tablenotemark{c}} 
& \colhead{U\tablenotemark{c}} & \colhead{UVW1\tablenotemark{c}} 
& \colhead{UVM2\tablenotemark{c}} & \colhead{UVW2\tablenotemark{c}}
}
\startdata
\multicolumn{10}{c}{\it April 2006}\\
\hline
$16-04$ & $400$ & $2.40\pm 0.09$ & $0.94$ & $1.36/41$ & $13.0$ & $13.4$ & $12.5$ & $12.3$ & $12.6$ & $12.6$\\
$26-04$ & $155$ & $2.4\pm 0.1$   & $1.31$ & $0.99/20$ & $12.8$ & $13.2$ & $12.3$ & $12.0$ & $12.4$ & $12.3$\\
\hline
\multicolumn{10}{c}{\it July-August 2006}\\
\hline
$29/31-07$ & $4916$ & $2.30\pm 0.03$, $1.19_{-0.11}^{+0.09}$, $2.80\pm 0.04$ & $3.43$ & $1.22/319$ & $12.6$ & $13.0$ & $12.1$ & $11.7$ & $12.0$ & $11.9$\\
$30/07$ & $3276$ & $2.25\pm 0.03$, $1.2\pm 0.1$, $2.81\pm 0.05$ & $3.61$ & $1.17/287$ & {} & {} & {} & {} & {} & {}\\
$01-08$ & $351$  & $2.62\pm 0.05$   & $2.90$ & $1.17/93$ & $12.5$ & $<12.8$ & $<12.0$ & $<11.3$ & $11.7$ & $11.7$\\
$02-08$ & $1842$ & $2.44_{-0.07}^{+0.05}$, $1.2\pm 0.2$, $2.90_{-0.08}^{+0.10}$ & $2.46$ & $1.06/195$ & $12.5$ & $12.9$ & $<12.0$ & $11.4$ & $11.8$ & $11.7$\\
$03-08$ & $1605$ & $2.24_{-0.13}^{+0.05}$, $1.1_{-0.2}^{+0.1}$, $2.75_{-0.08}^{+0.06}$ & $2.96$ & $1.31/205$ & $12.6$ & $12.9$ & $<12.0$ & $11.6$ & $11.9$ & $11.9$\\
$05-08$ & $517$  & $2.67\pm 0.05$   & $2.01$ & $0.96/93$ & $12.6$ & $13.0$ & $<12.0$ & $11.6$ & $11.9$ & $11.8$\\
$06-08$ & $295$  & $2.64\pm 0.08$   & $1.68$ & $0.91/52$ & $12.7$ & $13.0$ & $-$ & $11.6$ & $-$ & $-$\\
$08-08$ & $439$  & $2.62\pm 0.05$   & $2.29$ & $0.93/91$ & $-$ & $-$ & $-$ & $-$ & $-$ & $-$\\
$10-08$ & $318$  & $2.58\pm 0.06$   & $2.17$ & $0.91/70$ & $12.6$ & $12.9$ & $<12.0$ & $<11.3$ & $11.8$ & $11.7$\\
$12-08$ & $139$  & $2.8\pm 0.2$     & $1.27$ & $1.10/17$ & $12.5$ & $12.8$ & $<12.0$ & $<11.3$ & $12.2$ & $11.6$\\
$20-08$ & $184$  & $2.4\pm 0.1$     & $1.14$ & $1.09/25$ & $12.5$ & $12.9$ & $<12.0$ & $11.6$ & $11.9$ & $11.8$\\
$22-08$ & $161$  & $2.6\pm 0.1$     & $1.62$ & $1.42/25$ & $12.5$ & $12.9$ & $<12.0$ & $11.5$ & $11.9$ & $11.8$\\
\enddata
\tablenotetext{a}{$\Gamma$ for the power law model or $\Gamma_1$, $E_{\rm{break}}$ [keV], 
$\Gamma_2$, respectively, for the broken power law model. The absorption column is 
fixed to the Galactic value ($N_{\rm{H}}=1.36\times 10^{20}$~cm$^{-2}$, Lockman \& Savage 1995).}
\tablenotetext{b}{Observed flux in the $0.3-10$~keV band [$10^{-10}\times$~erg~cm$^{-2}$~s$^{-1}$].}
\tablenotetext{c}{Observed magnitudes. Error $0.1$ mag for all, including systematics. 
Lower limits indicate a saturation of the detector.}
\end{deluxetable}

\clearpage

\begin{table}
\begin{center}
\caption{Parameters for the SSC model by Ghisellini, Celotti \& Costamante (2002) 
used to interpolate the 
SED (Fig.~\ref{fig:sed}). $\Gamma_{\rm bulk}$ is the bulk Lorentz factor, $\theta$ 
the viewing angle, $\delta$ the Doppler factor, and $B$ the magnetic field. 
See the text for more details.\label{param}} 
\vskip 12pt
\begin{tabular}{llll}
\hline
            &Jul 29           &Aug 2     & Units \\
\hline
\hline
$R$                  & $5$               & $5$          & $10^{15}$~cm \\
$L^\prime_{\rm inj}$ & $1.1$             & $0.3$        & $10^{42}$~erg~s$^{-1}$  \\
$\gamma_{\rm break}$ & $1.5$             & $0.9$        & $10^4$\\
$\gamma_{\rm max}$   & $1.75$            & $1.1$        & $10^5$\\
$s$                  & $2.5$             & $2.6$        & \\
$B$                  & $0.27$            & $0.55$       & Gauss \\
$\Gamma_{\rm bulk}$  & $30$   		 & $30$         & \\
$\theta$             & $1.7$             & $1.7$        & degrees \\
$\delta$             & $33.5$            & $33.5$       &  \\
\hline
\end{tabular}
\end{center}
\end{table}

\clearpage

\clearpage

\clearpage

\begin{thebibliography}{}
\bibitem[2005]{hess1} Aharonian F., Akhperjanian A.G., Aye K.-M., et al., 2005a, A\&A 430, 865

\bibitem[2005]{hess3} Aharonian F., Akhperjanian A.G., Aye K.-M., et al., 2005b, A\&A 442, 895

\bibitem[Aharonian(2004)]{2004vhec.book.....A} Aharonian, F.~A.\ 2004, Very 
high energy cosmic gamma radiation : a crucial window on the extreme 
Universe, by F.A.~Aharonian.~River Edge, NJ: World Scientific Publishing, 
2004

\bibitem[2005]{BAT} Barthelmy S.D., Barbier L.M., Cummings J.R., et al., 2005, Space Sci. Rev. 120, 143

\bibitem[2006]{hess2} Benbow W., Costamante L., Giebels B. on behalf of the HESS Collaboration, 2006, ATel 867

\bibitem[2005]{XRT} Burrows D.N., Hill J.E., Nousek J.A., et al., 2005, Space Sci. Rev. 120, 165

\bibitem[2006]{RMFXRT} Campana S., Beardmore A.P., Cusumano G., Godet O., 2006, Swift XRT CALDB Release Note 09: 
Response matrices and Ancillary Response Files

\bibitem[1997]{catanese} Catanese M., Bradbury S.M., Breslin A.C., et al., 1997, ApJ 487, L143

\bibitem[1999]{chadwick} Chadwick P.M., Lyons K., T.J.L. McComb, et al., 1999, ApJ 513, 161

\bibitem[1999]{chiappetti} Chiappetti L., Maraschi L., Tavecchio F., et al., 1999, ApJ 521, 552

\bibitem[2006]{foschini1} Foschini L., Ghisellini G., Raiteri C.M., et al., 2006, A\&A 453, 829

\bibitem[2004]{gehrels} Gehrels N., Chincarini G., Giommi P., et al., 2004, ApJ 611, 1005

\bibitem[2002]{ghisellini1} Ghisellini G., Celotti A., Costamante L., 2002, A\&A 386, 833

\bibitem[2001]{kataoka} Kataoka J., Takahashi T., Wagner S.J., et al., 2001, ApJ 560, 659

\bibitem[1995]{lockman} Lockman F.J., Savage B.D., 1995, ApJSS 97, 1

\bibitem[Maraschi et al.(1999)]{1999ApJ...526L..81M} Maraschi, L., et al.\ 
1999, \apjl, 526, L81

\bibitem[1995]{padovani} Padovani P. \& Giommi P., 1995, ApJ 444, 567

\bibitem[1998]{elena} Pian E., Vacanti G., Tagliaferri G., et al., 1998, ApJ 492, L17

\bibitem[2006]{raue} Raue M. on behalf of the HESS Collaboration, 2006, In: The keV to TeV Connection. Roma, 17-19 October 2006 
[\texttt{http://gri.rm.iasf.cnr.it/keVtoTeV/Docs$\setminus$33\_Mraue.pdf}]

\bibitem[2005]{UVOT} Roming P.W.A., Kennedy T.E., Mason K.O., et al., 2005, Space Sci. Rev. 120, 95

\bibitem[1996]{stecker1} Stecker F.W., de Jager O.C., \& Salamon M.H., 1996, ApJ 473, L75

\bibitem[2006]{stecker2} Stecker F.W. \& Scully S.T., 2006, ApJ 652, L9

\bibitem[Takahashi et al.(2000)]{2000ApJ...542L.105T} Takahashi, T., et 
al.\ 2000, \apjl, 542, L105 

\bibitem[2001]{tanihata} Tanihata C., Urry C.M., Takahashi T., et al., 2001, ApJ 563, 569

\bibitem[1998]{fabrizio1} Tavecchio F., Maraschi L., \& Ghisellini G., 1998, ApJ 509, 608

\bibitem[2001]{fabrizio2} Tavecchio F., Maraschi L., Pian E., et al., 2001, ApJ 554, 725

\bibitem[1997]{urry} Urry C.M., Treves A., Maraschi L., et al., 1997, ApJ 486, 799

\bibitem[1995]{egret} Vestrand W.T., Stacy J.G., \& Sreekumar P., 1995, ApJ 454, L93

\bibitem[2005]{xuecui} Xue Y. \& Cui W., 2005, ApJ 622, 160

\bibitem[2002]{zhang} Zhang, Y.H., Treves, A., Celotti, A. et al., 2002, ApJ, 572, 762

\end{thebibliography}
\end{document}